\documentclass{svproc}
\usepackage{url}
\usepackage{amsmath}
\usepackage{graphicx}
\usepackage{caption}
\usepackage{float}
\floatstyle{boxed} 
\restylefloat{figure}

\graphicspath{ {./} }

\begin{document}
    \mainmatter              
    \title{Safeguarding National Security Interests Utilizing Location-Aware Camera Devices}
    \titlerunning{Applications of location-aware cameras}  
    %
    \author{Sreejith Gopinath{\inst{1}} \and Aspen Olmsted{\inst{2}}}
    \authorrunning{Gopinath and Olmsted et al.} 
    %
    \tocauthor{Sreejith Gopinath and Aspen Olmsted}
    \institute{NYU Tandon School of Engineering, New York, USA \\
    \and Simmons University, Boston, MA, USA \\
    \email{sree.gopinath@nyu.edu{\inst{1}} and olmsteda@simmons.edu{\inst{2}}}}

    \maketitle             

    \begin{abstract}
    The rapid advancement of technology has resulted in advanced camera capabilities coming to smaller form factors with improved energy efficiency. These improvements have led to more efficient and capable cameras on mobile devices like mobile phones, tablets, and even eyeglasses. Using these unobtrusive cameras, users can capture photographs and videos of almost any location where they have physical access. Unfortunately, the proliferation of highly compact cameras has threatened the privacy rights of individuals and even entire nations and governments. For example, governments may not want photographs or videos of sensitive installations or locations like airside operations of military bases or the inner areas of nuclear power plants to be captured for unapproved uses. In addition, solutions that obfuscate images in post-processing are subject to threats that could siphon unprocessed data. Our work proposes a Global Positioning System-based approach to restrict the ability of smart cameras to capture and store images of sensitive areas.
    \keywords{national security, photography, privacy, bounding boxes}
    \end{abstract}
    \section{Introduction}
    The topic of photographing the interior and exterior of federal facilities is indeed a touchy one. The Department of Homeland Security in the United States has promulgated Operational Readiness Order HQ-ORO-002-2018 \cite{dhs:op:readiness}, which lays out detailed guidelines and restrictions regarding the photography and videotaping of the interior and exterior of federal buildings. 18 United States Code 795 \cite{18:uscode:75} makes it a crime to photograph military installations without permission. In addition, 18 United States Code 797 \cite{18:uscode:797} makes it a crime to disseminate photographs of military bases without explicit permission. It is evident then that the restriction of photography in sensitive areas is a national security concern. Detailed images of sensitive sites easily accessible in the public domain aid bad actors in planning and executing physical attacks on the sites; yet, the proliferation of cameras with advanced features in hard-to-detect form factors and superior energy efficiency makes it increasingly difficult to enforce the laws and guidelines concerning the photography of such locations.

    In analyzing the \textbf{Ex}changeable \textbf{i}mage \textbf{f}ile (Exif) format data of a large corpus of image data available for public consumption on the internet, it is self-evident how a large amount of photographic data relates to areas that one might expect to restrict photography. One can easily find several images of sensitive airside areas of airports, military bases, aircraft maintenance facilities, air traffic control facilities, nuclear installations, and power plants. Photography is prohibited in several places and situations mentioned above. Notification of the prohibition is given through explicit notices or even incorporated into local laws. Unfortunately, the enforcement of the rule is mainly dependent on physical observation and prevention of the activity or confiscation of equipment. The proliferation of cameras and recording devices in small-form factors have exacerbated the problem. The increase stems from the difficulty of observing the restricted activity and preventing it from happening physically. As a result, the free dissemination of images of restricted areas poses threats to national security.

    While existing solutions rely on post-processing to maintain the privacy of the photographed subjects, the main thrust of this work is to propose a geographical bounding boxes-based solution that prevents devices from capturing images when they are located within sensitive geographical areas. When implemented correctly and on a global scale, this solution will help fortify the physical security postures of sites of national security importance. Section 2 presents relevant research that has been published in the past. Section 3 introduces solutions that have been proposed in the past for the same or closely related problems.  We also discuss in that section how our solution addresses the shortcomings of those solutions. Section 4 introduces our hypothesis with a few examples of use-cases our solution improves. Section 5 defines some of the terms and concepts used in our hypothesis. Section 6 describes the proposed evaluation that we will undertake and explains the various parameters used to evaluate our solution. Section 7 describes the multiple components of our solution and prototype in detail. Section 8 describes the implementation of our solution prototype in detail and describes the various workflows associated with numerous operations. Section 9 details our experience with evaluating the prototype thoroughly. Section 10 discusses the limitations and potential shortcomings of our proposed solution. Finally, section 11 concludes our work and offers a few extensions to work already been done.

    \section{Related Research}
    In their work on exploring privacy and security utilizing internet-connected cameras \cite{valente:junia:keerthi}, Valente et al. [2019] show how attackers can observe encrypted traffic to and from internet-connected cameras to observe the properties of what the camera is actually "seeing" at that time. They also discuss vulnerabilities in specific cameras (and other cameras utilizing the same software or hardware platforms) that attackers may use to invade privacy. This work underscores the premise that it is easy for sensitive photographs to fall into the wrong hands if they are in the public domain.

    In their work on exploring ways to protect privacy in video graphic data captured by self-driving car cameras \cite{xiong:cai:han:alrawais}, Xiong et al. [2021] propose methods to prevent the invasion of individuals' privacy. They suggest obfuscating the data in post-processing steps while retaining major, essential features. However, since the obfuscation happens during post-processing, there is still a risk of the actual, unprocessed photographic data being leaked. Furthermore, since the significant parts of the data are still retained, this method might not work well to protect the privacy of substantial installations of national security importance, such as military bases and nuclear power plants. 

    In their work discussing various \textbf{I}nternet \textbf{P}rotocol (IP) cameras that store captured data in the cloud \cite{liranzo:hayajneh}, Liranzo, J et al. [2017] describe various vulnerabilities and possible attack vectors that threaten to invade users' privacy by illegally accessing the image and video data captured by these cameras. The authors also preview recommendations on securing the data by employing encryption both in-flight and at rest and generally regulating access to the stored data in the cloud. The risks called out by this work prove that it is not just enough to have a post-processing step to obfuscate sensitive data because the device may be compromised before the obfuscation is completed.

    In their paper describing a storage system for video data \cite{yu:chen:wu:cai:cui}, J. Yu et al. [2020] describe a low-cost distributed storage system architecture for video data to fend off geo-range attacks. In their work on discussing solutions to privacy concerns in wear-able cameras \cite{hassan:sazonov}, M. A. Hassan et al. [2020] discuss how privacy concerns may creep into the usage of egocentric wearable cameras. They propose using deep learning neural networks to apply image redaction by selective removal to address the privacy concerns of bystanders or other parties. This work also relies on post-processing to attain privacy objectives (high computational costs).

    The general risk with solutions that propose post-processing to make data unappealing is that the data that has already been collected could fall into the wrong hands before post-processing renders it useless.

    In their doctoral dissertation \cite{jiayu:phd:2019}, Shu J discusses privacy issues relating to using cameras in a social setting. They describe an architecture whereby users of photo and video cameras are informed of individuals' privacy preferences close to the devices; they also propose a method of visual gestures that allows individuals to make their privacy preferences explicitly known. The method proposed in this dissertation employs tags and visible gestures, but it could prove exhausting to individuals trying to protect their privacy. Furthermore, this method does not address how the confidentiality requirements of installations like military bases and nuclear power plants can be enforced.

    In their paper addressing privacy concerns in smart cities, \cite{peters:hanvey:veluru:mady:boubekeur:nuseibeh} F. Peters et al. [2018] propose setting up privacy zones in smart cities. The framework makes sure that individuals are aware of privacy factors before sharing their data in these privacy zones and recommends actions that individuals can take to safeguard their privacy. However, the drawback of this solution is that it does not prevent the deliberate exfiltration of sensitive photographic data for unauthorized use.

    In their paper on exploring solutions to data sharing and privacy \cite{koufogiannis:pappas}, F. Koufogiannis et al. [2016] hypothesize the usage of adding noise to data to preserve users' privacy when they share data. They hypothesize that adding less noise is feasible where there is a high physical concentration of users in a given geographical location and inversely in an area of lower concentration. This paper deals solely with the privacy concerns of sharing location data and does not deal with the risks of misusing sensitive photographs.

    \section{Other solutions proposed in the past}
    In this section, we discuss some solutions that have been attempted in the past. We also discuss some shortcomings of these proposals that do not fully address the privacy concerns we are researching.
        \subsection{Neural networks}
        Generative Adversarial Networks have been proposed to post-process video data and obfuscate non-essential features in the images to ensure that someone cannot infer the physical location of the self-driving car from the video graphics data. A similar approach has been suggested to remove bystander information potentially invading another person's privacy from egocentric wearable cameras. This solution involves post-processing photographic and video data to remove content following privacy restrictions. 

        As shown in the previous section, the drawback of this approach is that the data is not safe in the interim between collection and completion and post-processing. If post-processing does not replace the original dataset, the risk of the original dataset falling into the wrong hands is perpetually existent. This process is also computationally costly and does not preclude capturing restricted photographs; it only serves to try and prevent the dissemination of these photographs.
        \subsection{Recommended best practices}
        \begin{itemize}
            \item Purchase photo and video equipment only from reputed vendors that have a strong track record of keeping users' privacy at the forefront.
            \item Users update the software/device firmware regularly and apply all available software security patches.
            \item Manufacturers to put safeguards against brute force attacks that attempt to guess passwords.
            \item Implement on-device cryptography to keep data safe.
            \item Use frameworks that use \textbf{G}lobal \textbf{P}ositioning \textbf{S}ystem (GPS) data to alert users to infer the users' location  from shared data.
            \item Scrub location and other identifying data from Exif before disseminating photographic data.
        \end{itemize}

        These recommendations provide a potential path to preventing sensitive photographs and video captured by the devices from being unknowingly shared with other parties. Recommendations also spell out how the privacy of the users of these devices may be protected. However, recommendations are just advisory and do not serve to safeguard the confidentiality of the entities that are being photographed.
        \subsection{Visual cues}
        \begin{itemize}
            \item A beacon-based system alerts individuals close to the device that photo or video information is collected.
            \item Visual gestures or tags inform the user/device that there are privacy concerns and that the device should cease recording.
        \end{itemize}

        These devices are commonly found in places that display hazardous objects that should not be approached. Beacons that warn about prevailing photography restrictions in a given area are mainly advisory and do not have enforcement capabilities, unlike, for example, cell phone jammers.
        \subsection{Adding noise to shared data}
        Based on GPS data, determine whether an individual user is in a highly-populated area and add more noise to the shared data.

        This solution also deals with post-processing photo and video data after being collected. In addition to being computationally expensive, this solution relies on post-processing efficiency to ensure that private data is not leaked to unauthorized consumers.

        In our study of the literature, the solutions proposed so far fall into two main classifications:
        \begin{itemize}
            \item Solutions that are advisory; and
            \item Solutions that attempt to post-process the data to remove sensitive data.
        \end{itemize}
        
        Solutions that are advisory do not prevent the wilful collection and usage of sensitive photographs. In addition, solutions that propose post-processing techniques to obfuscate all or part of the data suffer from drawbacks in that devices could be compromised, and photographic data stolen before the post-processing step. To overcome the shortcomings identified in either class of solutions discussed above, we propose a solution that prevents photographs of sensitive sites from being captured in the first place.

    \section{Hypothesis}
    We propose a solution that prevents photographs of sensitive sites from being captured in the first place. The prevention renders moot the questions around safeguarding the data before post-processing has taken place and whether it may be possible to reconstruct the original data from the post-processed data using advanced techniques. The bedrock of our proposal is a global database of geographical locations. Geographical locations are bounding boxes, defined as an area bounded by two latitudes and two longitudes. The database is treated as an exclude list which means that any bounding box present in this database represents a geographic location in which photography and videotaping are restricted. We are proposing the setup of a database containing a list of geographic bounding boxes representing areas or properties owned by governments (airports, nuclear installations, military bases, power plants, etc.). Governments or other authorities that designate geographic sites as sensitive will create bounding boxes using geographical coordinates and persist them in the global database. We propose database schemas that persist the restricted bounding boxes into a central database and a lightweight database on the device itself that acts as a cache. 

    We propose that the functionality to restrict or allow photography based on geographical location be built into the firmware of the camera and its control module. The device will need internet connectivity occasionally to initialize usage in each country or province that it is being used. The network connectivity gives the device a chance to download all restricted bounding boxes within that geographical region before being used for the first time. Devices will check in with the service at intervals based on changes in the GPS location of the device.

    We further propose that national governments mandate the use of this framework in camera devices that are sold or imported into their countries. As the adoption of this framework spreads, we believe that it will reduce the problem of unregulated photography to levels where physical observation and prohibition will become feasible again to stop recalcitrant individuals.
    
    \section{Definition of Terms}
        \subsection{Global Positioning System}
        The Global Positioning System (GPS) is a navigation system that uses satellites, receivers, and algorithms to determine an object's location, direction of travel, and velocity on earth or in the air. Three satellites help determine the position of an object by triangulation, i.e., by plotting the receiver's distance from each of the receivers. A fourth satellite is used to verify the information from the three satellites used in triangulation and to aid in deducing the latitude information of an object.
        \subsection{Bounding Boxes}
        A bounding box is a geographical area defined by two latitudes and two longitudes. The conventional format of a bounding box is
        \begin{equation}
            \begin{split}
                bounding\_box = [min(longitude_1, longitude_2),min(latitude_1, latitude_2),\\max(longitude_1, longitude_2),max(latitude_1, latitude_2)]
            \end{split}
        \end{equation}

        \subsection{Centroid of a Bounding Box}
        The centroid of a geographical bounding box is defined as the intersection point of the two diagonals. The coordinates of the centroid of the bounding box are calculated as
        \begin{equation}
            centroid = \dfrac{(longitude_1 + longitude_2)}{2} , \dfrac{(latitude_1 + latitude_2)}{2}
        \end{equation}

    \section{Proposed evaluation}
    Since we cannot modify the camera firmware on an iPhone, we plan to build a mobile camera application to demonstrate the feasibility of our idea. Our mobile application will be developed on the iOS platform and has basic capabilities such as capturing a photograph and saving it to device storage. In addition, the application has a small local database to hold a subset of locally restricted boundary boxes. 

    The application fetches restricted bounding boxes in the country or province of usage. As we see in later sections, the storage costs of storing a single bounding box are low. Therefore it is feasible to keep bounding boxes for the entire country on the device at a given time. If the bounding box density in a particular country is prohibitively high, the framework can split the country into manageable regions in a way that is opaque to the device. However, when the device detects that it is currently located in a region for which it has not downloaded bounding boxes, it prevents the camera from functioning until internet connectivity has been restored and bounding boxes have been downloaded. In addition, the framework shall allow the pre-downloading of bounding boxes for regions in which travel is anticipated so that the device can work on arrival in those regions.

    The application has a module that records just one of the GPS locations of the device. When the application is active, it checks the GPS location of the device once every ten minutes. The application only stores the device's most recent location to preserve privacy. If the device has traveled more than a mile from its last recorded location, the module calls out to the service and fetches more restricted bounding boxes in the new vicinity of the device. Each device is to be configured with a "permissible distance" based on the optical capabilities of the camera lenses. If the device is within the permissible distance of a restricted bounding box, the camera application is not allowed to capture images. The concept of "permissible distance" prevents users from working around the restriction by situating the device outside the bounding box and using telephoto capabilities to photograph the area within a restricted bounding box.

    \section{Description of the prototype}
    We have built out all the modules of the solution. The solution has hardware and software components, and they are described in the subsections below.
    \subsection{Hardware platforms}
        \subsubsection{Database system}~\\
        We installed a Postgres database on a Raspberry Pi 4. The Raspberry Pi computer has 8GB synchronous dynamic random access memory (SDRAM).

        We use a SQLite database to store the restricted bounding boxes locally on the device for the application cache.
        \subsubsection{The photography application}~\\
        We have built a photography application whose only function is to capture pictures and save them to the photo Gallery. We tested the application on an iPhone 13 Pro.
    \subsection{Software Components}
        \subsubsection{iOS application to capture images}~\\
        We developed an iOS application that utilizes the camera of the iPhone to take pictures and save them to the gallery.
        \subsubsection{Database schema to hold restricted bounding boxes}~\\
        We created a database that contains the list of geographic bounding boxes representing areas where photography is restricted.
        \subsubsection{Application Programming Interface (API) framework to facilitate interaction with the database}~\\
        We implemented a \textbf{Re}presentational \textbf{S}tate \textbf{T}ransfer (ReST) API service using Spring Boot. This API has the following capabilities:
        \begin{itemize}
            \item POST~\\
            Used to add new bounding boxes to the database.
            \item GET~\\
            Returns a list of bounding boxes within a defined radius surrounding the GPS location passed in as an argument to the API.
        \end{itemize}
    
    \section{Implementation of the Prototype}
    \begin{figure}[htb]
        \centering
        \includegraphics[scale=0.2]{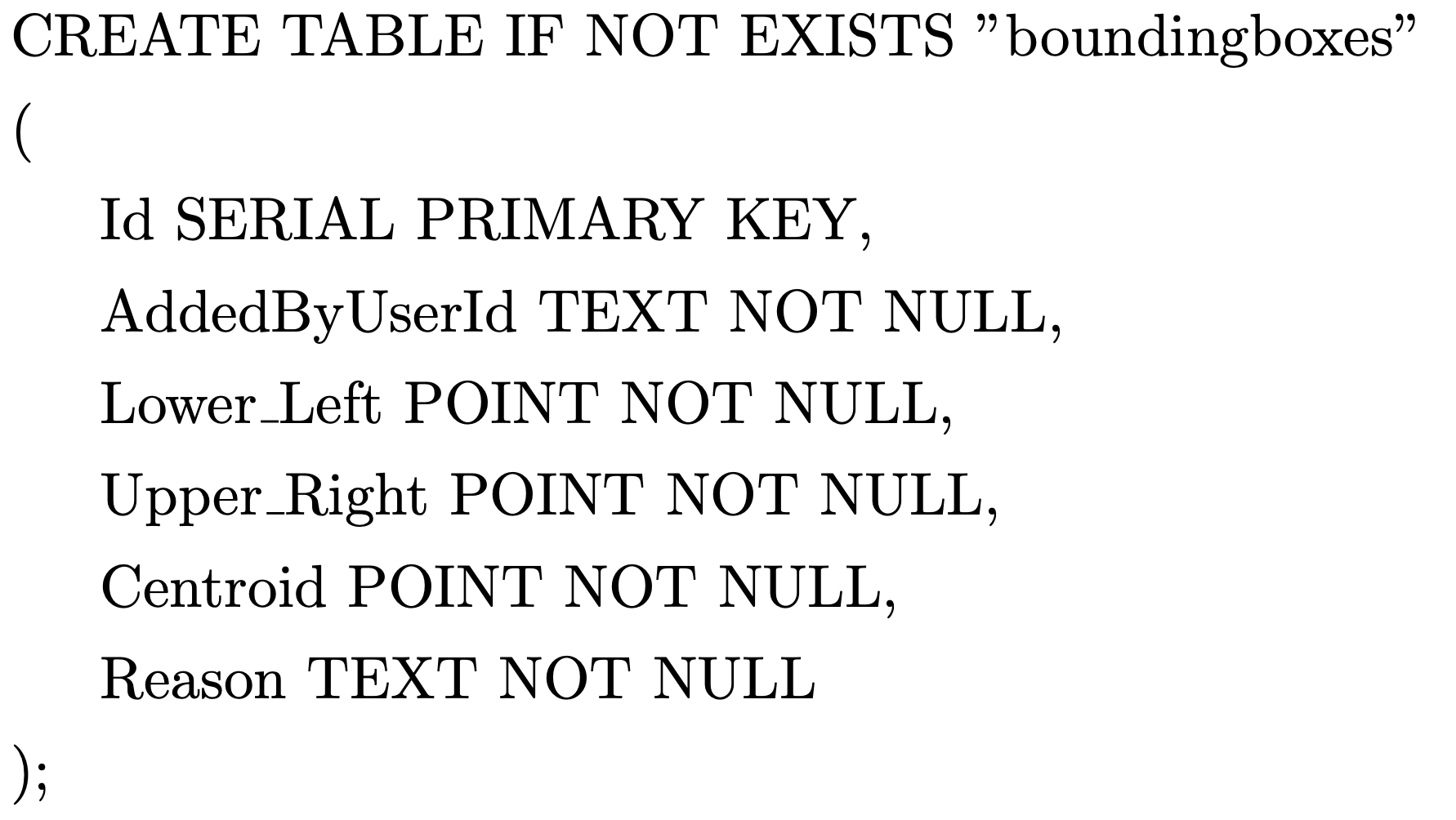}
        \caption{Schema of bounding box table}
        \label{fig:bbox_tab_schema}
    \end{figure}
    \subsection{Database of restricted bounding boxes}
    The schema of the main table is shown in Fig. \ref{fig:bbox_tab_schema}. Each bounding box in the database table gets its unique identifier. The bounding box is defined by two sets of latitudes and longitudes per convention. For audit purposes, we record the handle of the user or entity that added the bounding box to the database and the reason behind adding the restriction. As the bounding box is being added, the processing engine also computes the centroid of the bounding box and persists it along with the data. A centroid to compute against makes the whole process more efficient when the bounding boxes are accessed later to check for proximity.
    
    \subsection{ReST API Framework}
    The ReST API framework was developed as a Spring Boot application in Java. The API has two endpoints, as enumerated below.
        \subsubsection{Add a new restricting bounding box}
            \begin{figure}[htb]
                \centering
                \includegraphics[scale=0.3]{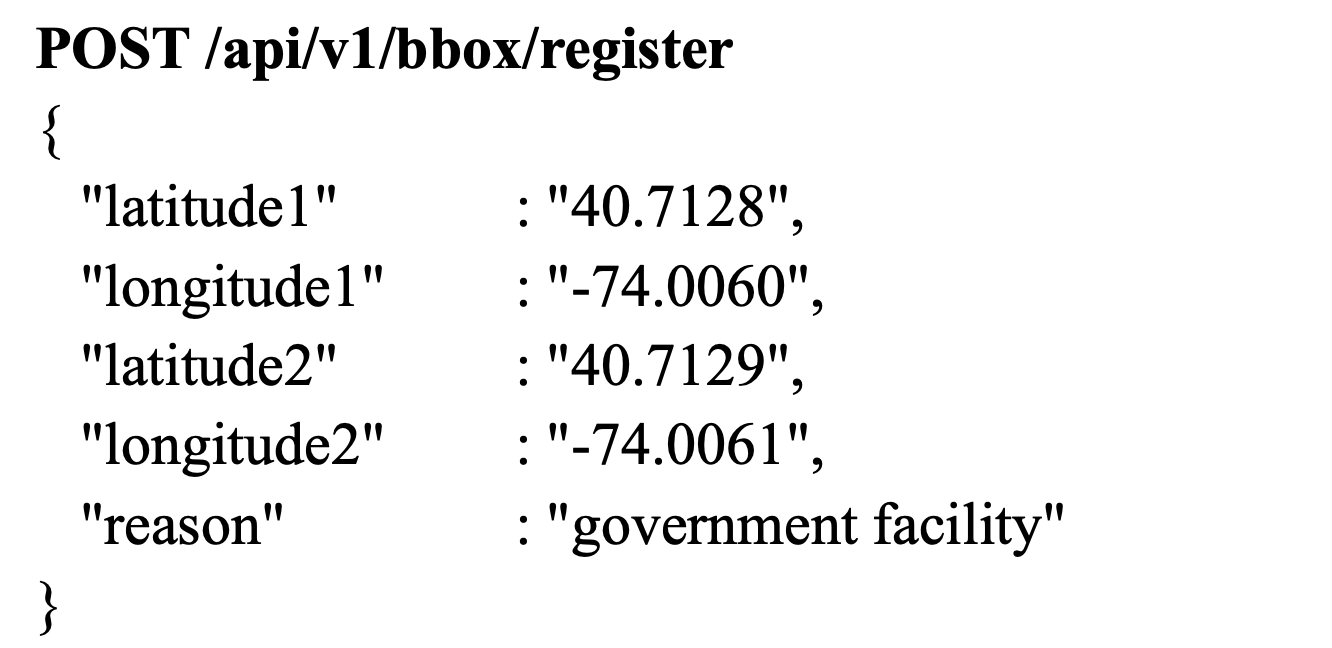}
                \caption{Request to add a new bounding box}
                \label{fig:regist_bbox}
            \end{figure}~\\
    The endpoint specification to add a new restricted bounding box is shown in Fig. \ref{fig:regist_bbox}. The endpoint above accepts a POST request to add a restricted bounding box into the database. A bounding box is defined by two points in two-dimensional space, the first representing the lower-left corner of the box and the other representing the upper-right corner of the box. The endpoint responds with a standard \textbf{H}yper\textbf{t}ext \textbf{T}ransfer \textbf{P}rotocol (HTTP)  status code denoting the result of the operation.
    \subsubsection{Retrieve a list of bounding boxes in the vicinity of the device’s current location}
    \begin{figure}[htb]
        \centering
        \includegraphics[scale=0.2]{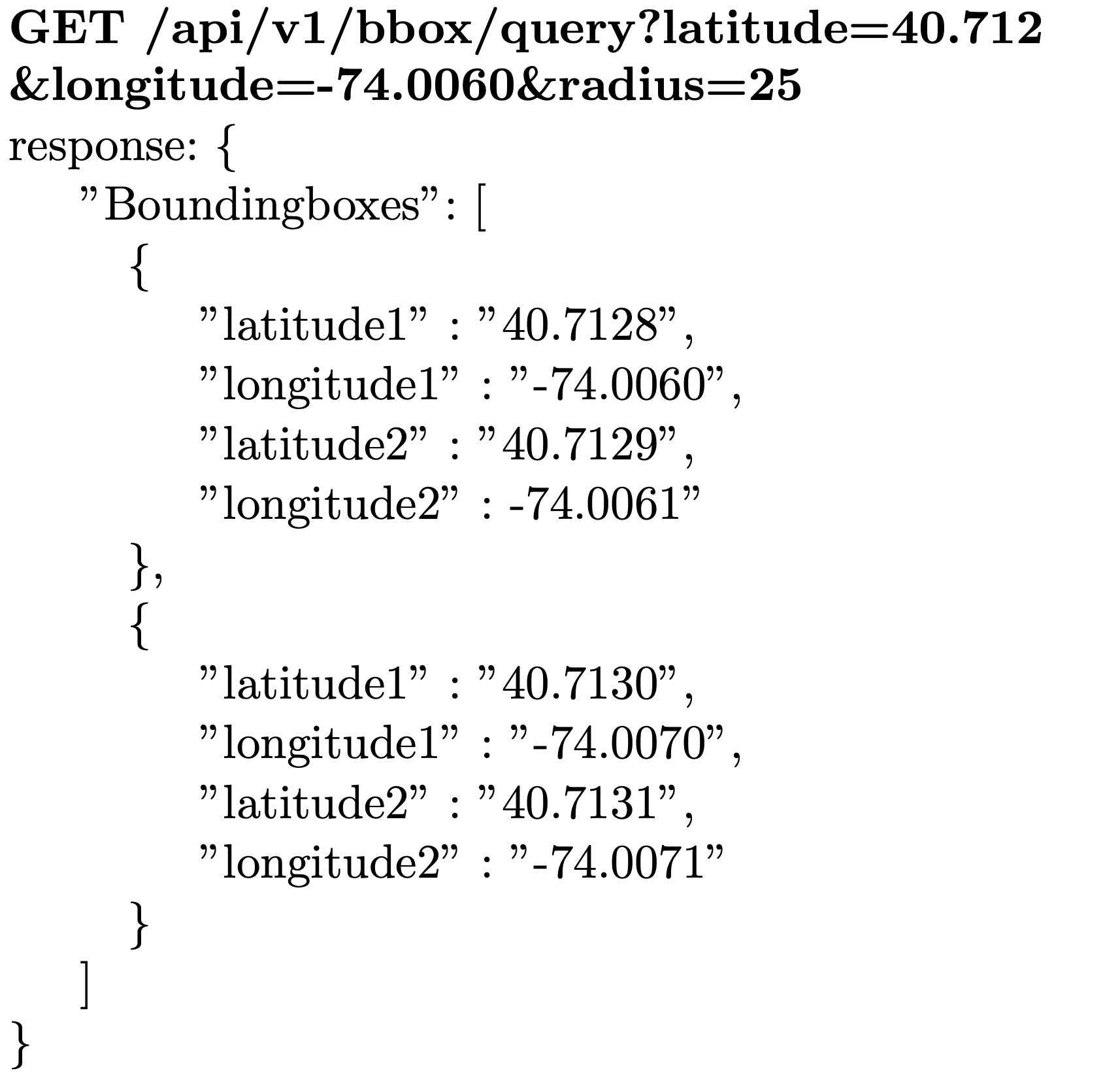}
        \caption{Request to fetch bounding boxes}
        \label{fig:api_fetch_bbox}
    \end{figure}
    ~\\Fig. \ref{fig:api_fetch_bbox} shows that the endpoint accepts a GET request that specifies the coordinates of the current location of the device as well as the desired radius of the vicinity in which to search for restricted bounding boxes. The endpoint responds with an array of bounding boxes that lie within the requested area of the device, along with the standard HTTP status codes denoting the result of the operation.
    \subsection{Processing Engine}
    The processing engine is the software component that translates the user input received via the API framework into actions on the database. 

    When the device requests all existing bounding boxes that exist in the current vicinity of the device, the processing engine uses the Haversine formula to compute the great circle distance between the current location of the device \((latitude_1, longtitude_1\)) and the centroid of each bounding box in the database \((latitude_2, longitude_2\)). If the great circle distance so computed is within the radius specified by the device in the request, the bounding box is added to the list returned to the device to be added to its local cache.

    The haversine formula to compute the great circle distance between any two points on earth is given as
    \[ a = sin^2(\dfrac{\Delta \varphi}{2}) + cos(\varphi_1)\cdot cos(\varphi_2) \cdot sin^2(\dfrac{\Delta \lambda}{2})\]
    \[ c = 2 \cdot atan2(\sqrt{a}, \sqrt{(1-a}) \]
    \[ distance = (radius\ of\ the\ earth\ in\ metres)\cdot c\]
    \[where\ \varphi\ is\ the\ latitude\ and\ \lambda\ is\ the\ longitude.\]

    \subsection{Database of restricted bounding boxes on device memory}
    \begin{figure}[htb]
        \centering
        \includegraphics[scale=0.2]{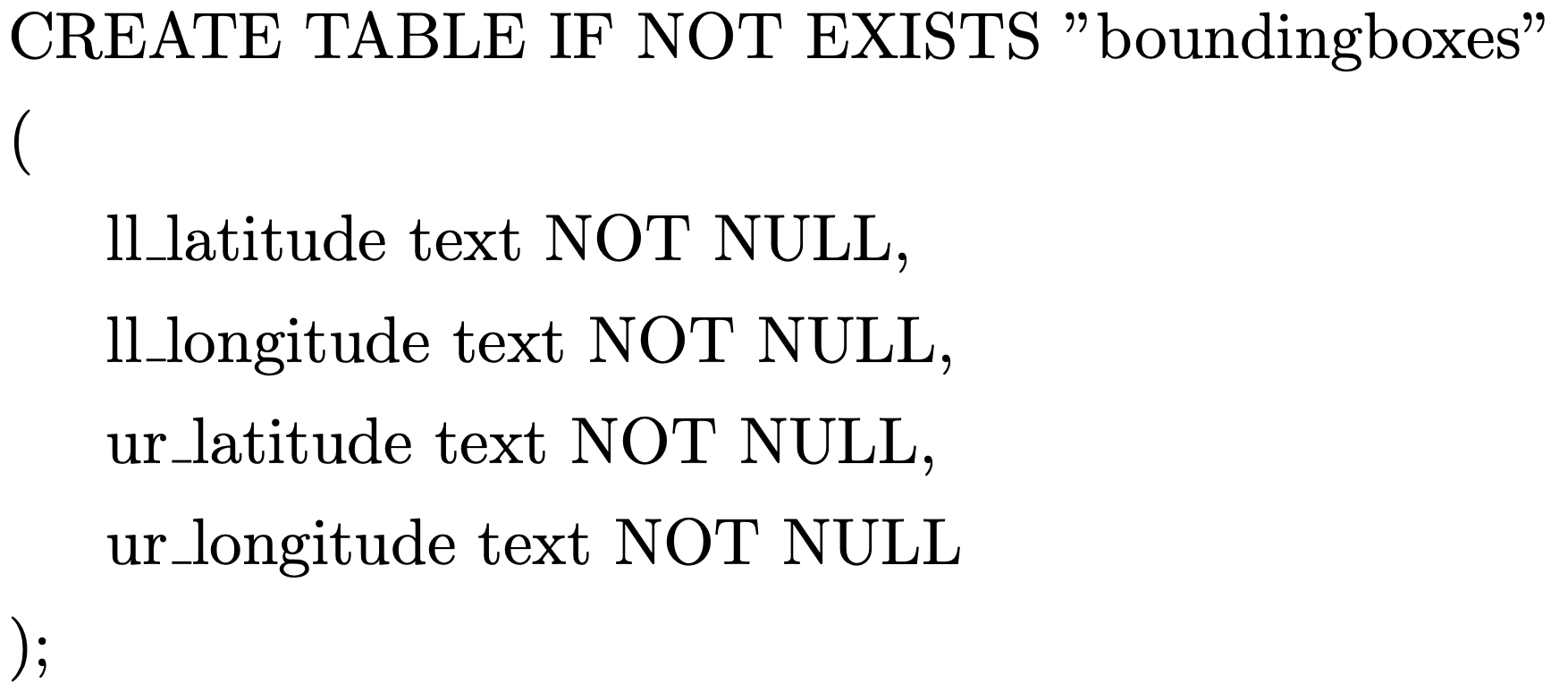}
        \caption{Schema of bounding boxes table on device memory}
        \label{fig:bbox_schema_device}
    \end{figure}
    ~\\Fig. \ref{fig:bbox_schema_device} shows the schema of the bounding boxes table on device memory.
    \subsection{Workflows for common operations}
        \subsubsection{Add a new bounding box}
\begin{figure}[htb]
            \centering
            \includegraphics[scale=0.2]{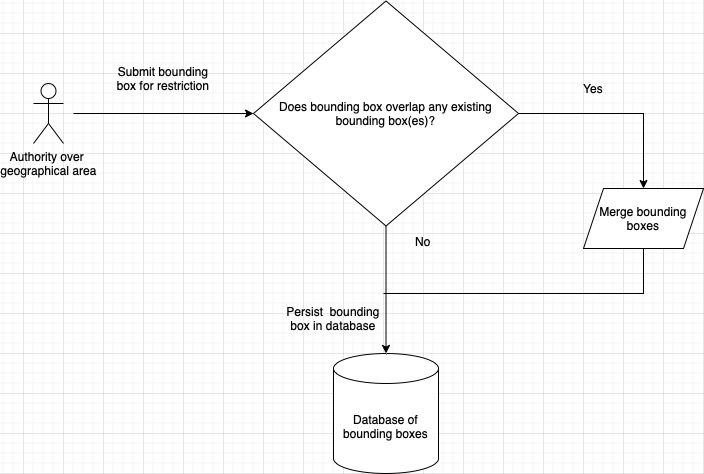}
            \caption{New bounding box insertion workflow}
            \label{fig:bbox_insert_wflow}
        \end{figure}

~\\Fig. \ref{fig:bbox_insert_wflow} shows the workflow when a new restricted bounding box is added to the database. A user with authority over a geographical area adds a restricted bounding box to the database via the API framework. Fig. \ref{fig:bbox_insert_wflow} shows how the processing engine processes a request to add a restrictive bounding box to the database. A group of nearby bounding boxes is retrieved from the database. Next, the engine runs a matching algorithm that checks whether the new bounding box overlaps any existing bounding boxes. If overlaps are detected, the engine adjusts them into discrete non-overlapping bounding boxes and persists them back into the database.
    \begin{figure}[htb]
        \centering
        \includegraphics[scale=0.2]{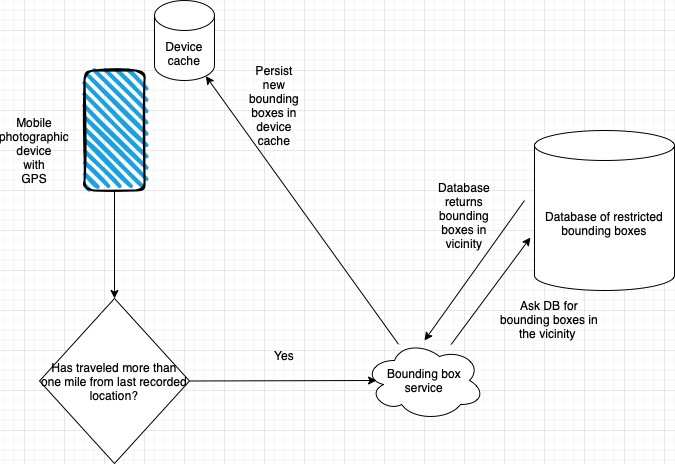}
        \caption{Bounding boxes refresh workflow}
        \label{fig:bbox_refresh_wflow}
    \end{figure}
    \subsubsection{Device refreshing its restricted bounding boxes cache}~\\
    Fig. \ref{fig:bbox_refresh_wflow} shows the workflow when a device needs to restore its local store of limited bounding boxes.
    \begin{figure}[htb]
        \centering
        \includegraphics[scale=0.2]{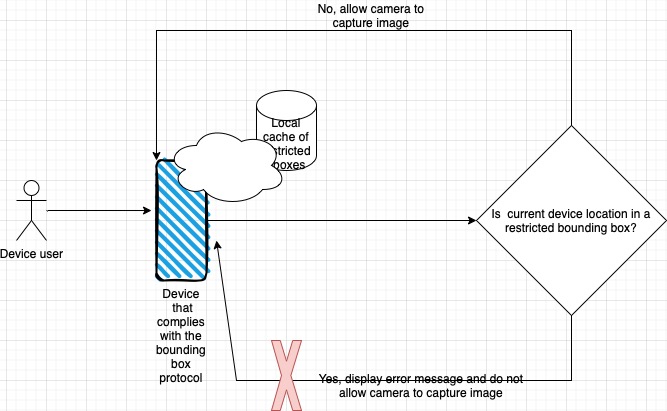}
        \caption{Image capture workflow}
        \label{fig:image_capt_wflow}
    \end{figure}
    As seen in Fig. \ref{fig:image_capt_wflow}, when the application starts up for the first time, it compares the current location of the device to the last recorded GPS location and informs the user that it needs to:
    \begin{enumerate}
        \item Record the GPS location of the device at ten-minute intervals when the application is active. Only the most recent GPS location is stored; GPS trails are not stored on the device to protect privacy.
        \item Connect to the internet to refresh the list of restricted bounding boxes in the national or provincial vicinity of the device.
    \end{enumerate}
    
    The device connects to the internet to download the list of restricted bounding boxes in the current vicinity of the device. As long as the camera application is active, it records the GPS location of the device and checks whether it has traveled more than a mile from its previously registered location. Suppose it has; the application calls the bounding box service to download the restricted bounding boxes within a 25-mile radius of the device's current location. Again, this continues for as long as the camera application on the device is active. The call to the service is an attempt to pick up any new bounding boxes added since the list was last updated.

    In addition, time triggers are built into the software. For example, if more than 24 hours have passed since the bounding box list was last updated, the device attempts to establish internet connection and refresh the inventory. At the 30-day mark, the camera on the device becomes inoperable until the refresh has been performed.

    \subsubsection{Device capturing an image}~\\
    Fig. \ref{fig:image_capt_wflow} depicts the device's workflow when it needs to take a picture.
    When the application receives a request to take a picture (via voice commands or tactile input by the user), it performs the following checks as depicted in figure \ref{fig:bbox_refresh_wflow}:
    \begin{enumerate}
        \item Has the device refreshed its local bounding box cache within the last configured positional and temporal triggers? If not, refresh the local cache before displaying the controls to capture the image.
        \item Does the device's current location lie within a permissible distance of any of the restricted bounding boxes on the device's local cache?
    \end{enumerate}

    If the current location does lie inside one of the limited bounding boxes, the device presents an error message to the user, and the picture is not captured.

    If the current location does not lie inside one of the restricted bounding boxes, the action to capture an image and persist it in the image gallery is completed.

    \section{Evaluation of the prototype}
    \subsection{Functional evaluation}
        \subsubsection{Functional evaluation of the server-side components.}
        \begin{enumerate}
            \item Adding non-overlapping restricted bounding boxes.~\\
            We used a free map application to draw bounding boxes and record the geographical coordinates. We then run two test cases to test the logic in the processing engine.
        \begin{enumerate}
            \item For the first test, we added two non-overlapping bounding boxes near each other.
            \item We added a third bounding box for the second test that intersected both existing bounding boxes.
        \end{enumerate} 
        \begin{figure}[htb]
            \centering
            \includegraphics[scale=0.2]{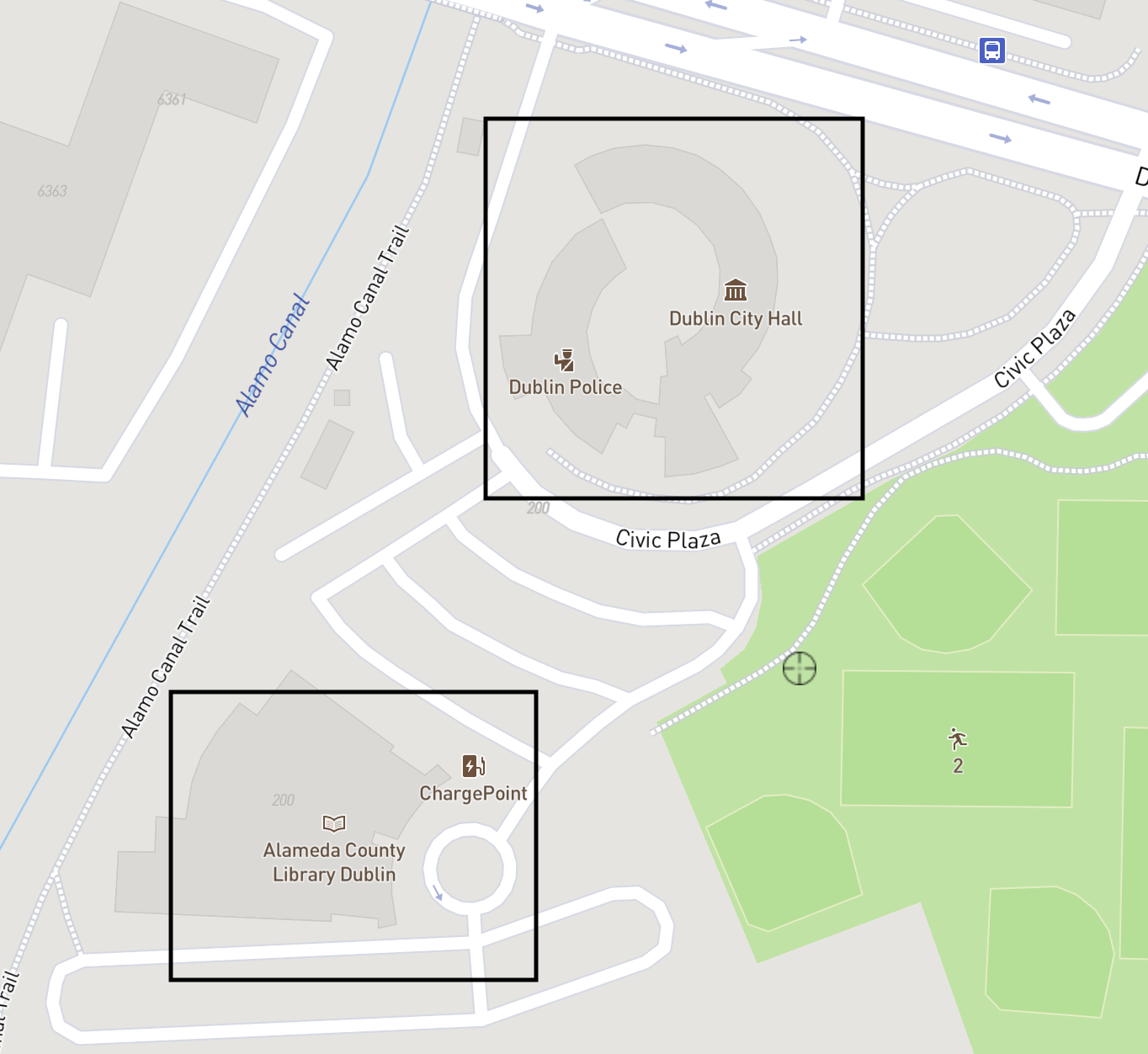}
            \caption{Visual representation of the non-overlapping bounding boxes that were added}
            \label{fig:non_overl_bboxes}
        \end{figure}
        \begin{figure}[htb]
            \centering
            \includegraphics[scale=0.2]{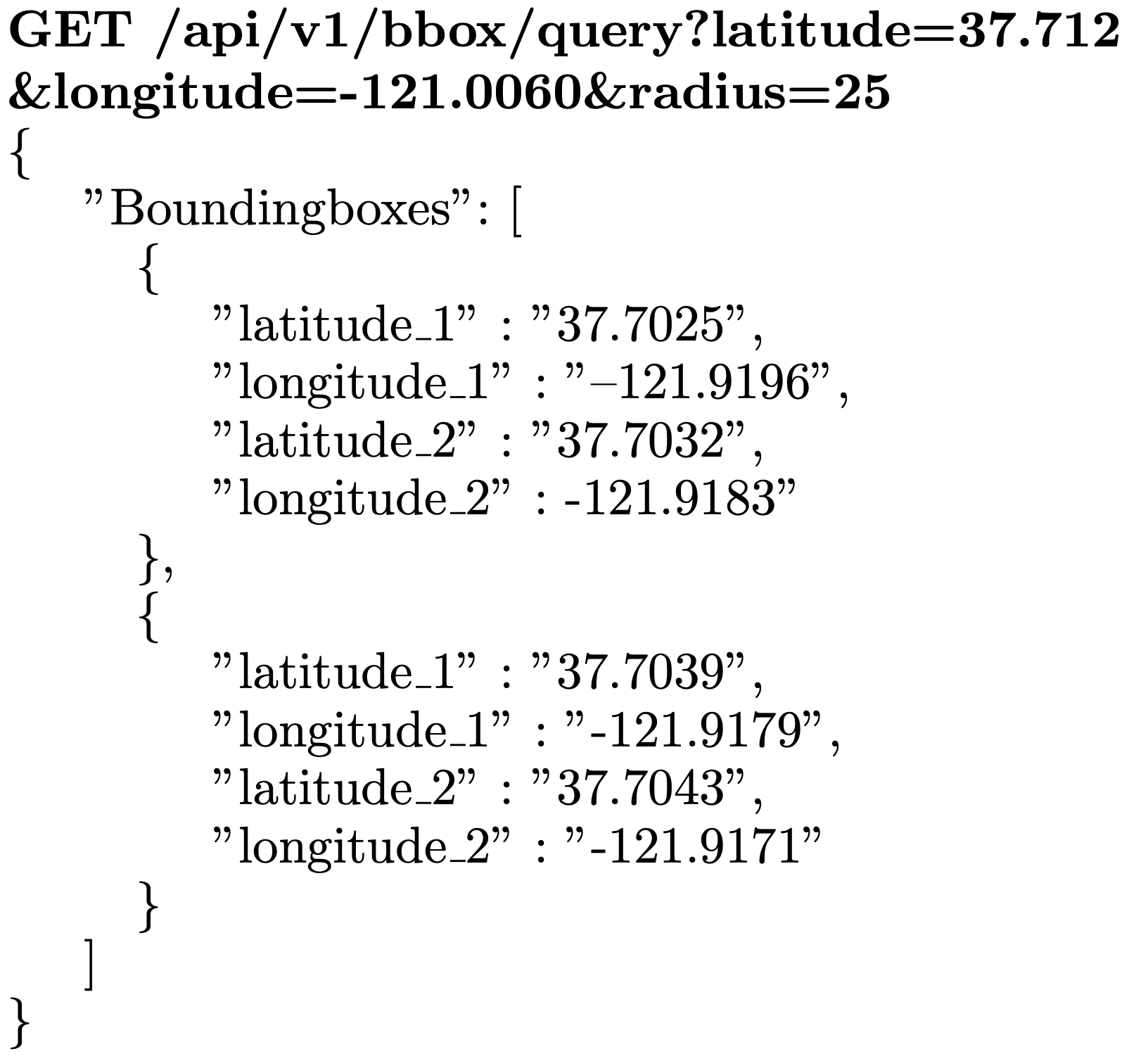}
            \caption{Fetch bounding boxes in the vicinity of the device}
            \label{fig:fetch_bboxes_vicinity}
        \end{figure} 
        \begin{figure}[htb]
            \centering
            \includegraphics[scale=0.2]{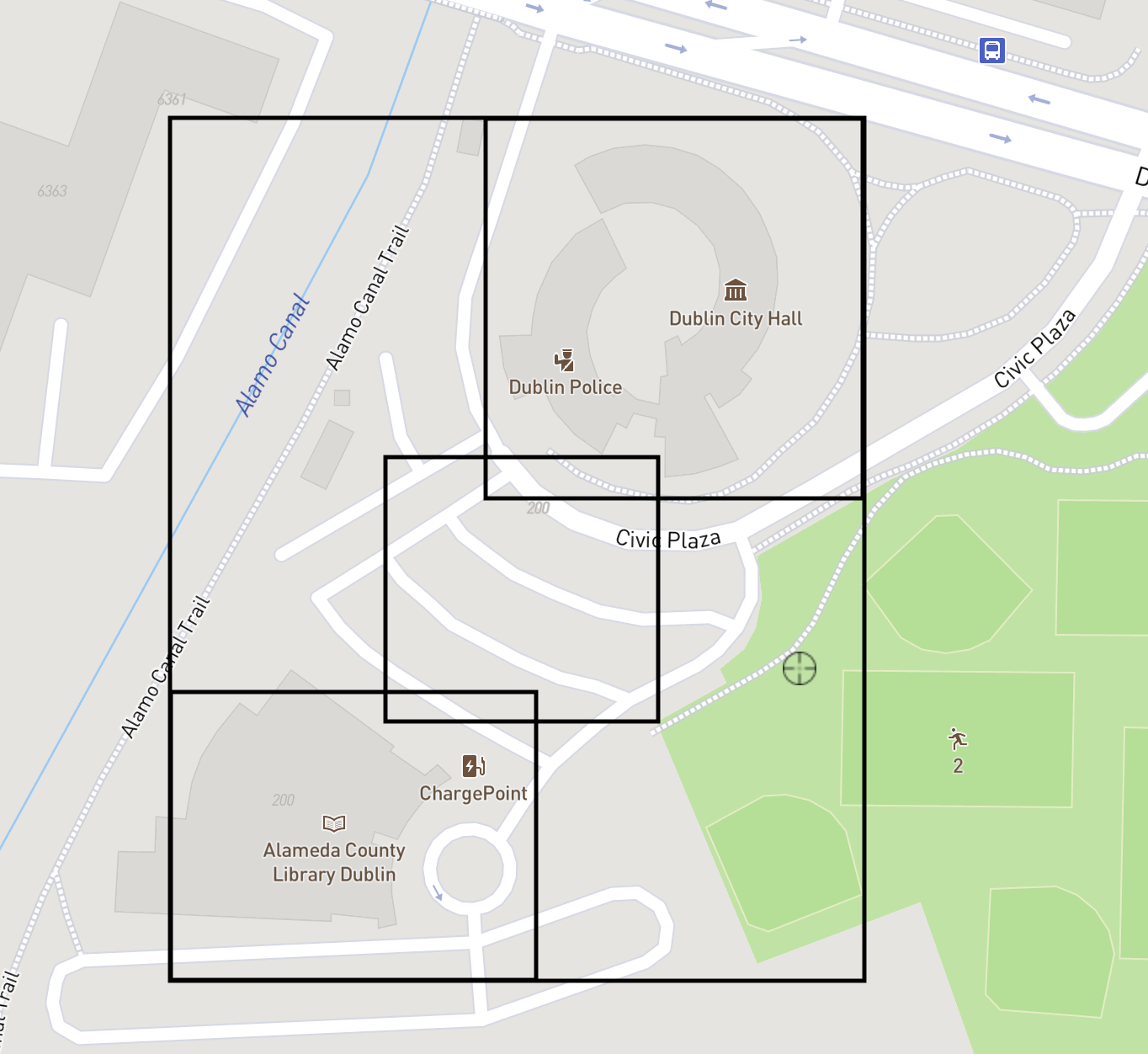}
            \caption{Visual representation of the encompassing bounding box when there are overlaps}
            \label{fig:encompass_bboxes}
        \end{figure}
        
        Fig. \ref{fig:non_overl_bboxes} shows the geographical bounding boxes as plotted on a map. When the application queried the database after these two bounding boxes had been added, it returned the two bounding boxes separately, as expected.
        
        Fig. \ref{fig:fetch_bboxes_vicinity} shows the API response when fetching bounding boxes in the vicinity of the device.
        \item Adding a bounding box that intersects existing restricted bounding boxes.
        
        Fig. \ref{fig:encompass_bboxes} shows the overlapping bounding boxes as plotted on a map. When the application queried the restricted bounding boxes after this, the processing engine worked correctly to return the correct all-encompassing bounding box.
        \end{enumerate}
        
        \subsubsection{Functional testing of the camera application.}
        \begin{enumerate}
            \item Testing the application on a fresh start.\\

            We connected the Raspberry Pi hosting the database to a portable battery unit and linked the device and the Raspberry Pi on the local hotspot network over Wifi. We then traveled to the vicinity of a restricted bounding box with the device turned on.  We intentionally left the application unstarted for this step. When launched, the refresh module contacted the API service running on the Raspberry Pi and fetched the bounding boxes in the current country of the device. When the device was outside any of the bounding boxes, the application allowed pictures to be captured. However, when the device had entered into a restricted bounding box, the application prevented the camera from capturing an image.\\

            \item Testing the application while it was active and the device was traveling.

            \begin{figure}[htb]
                \centering
                \includegraphics[scale=0.2]{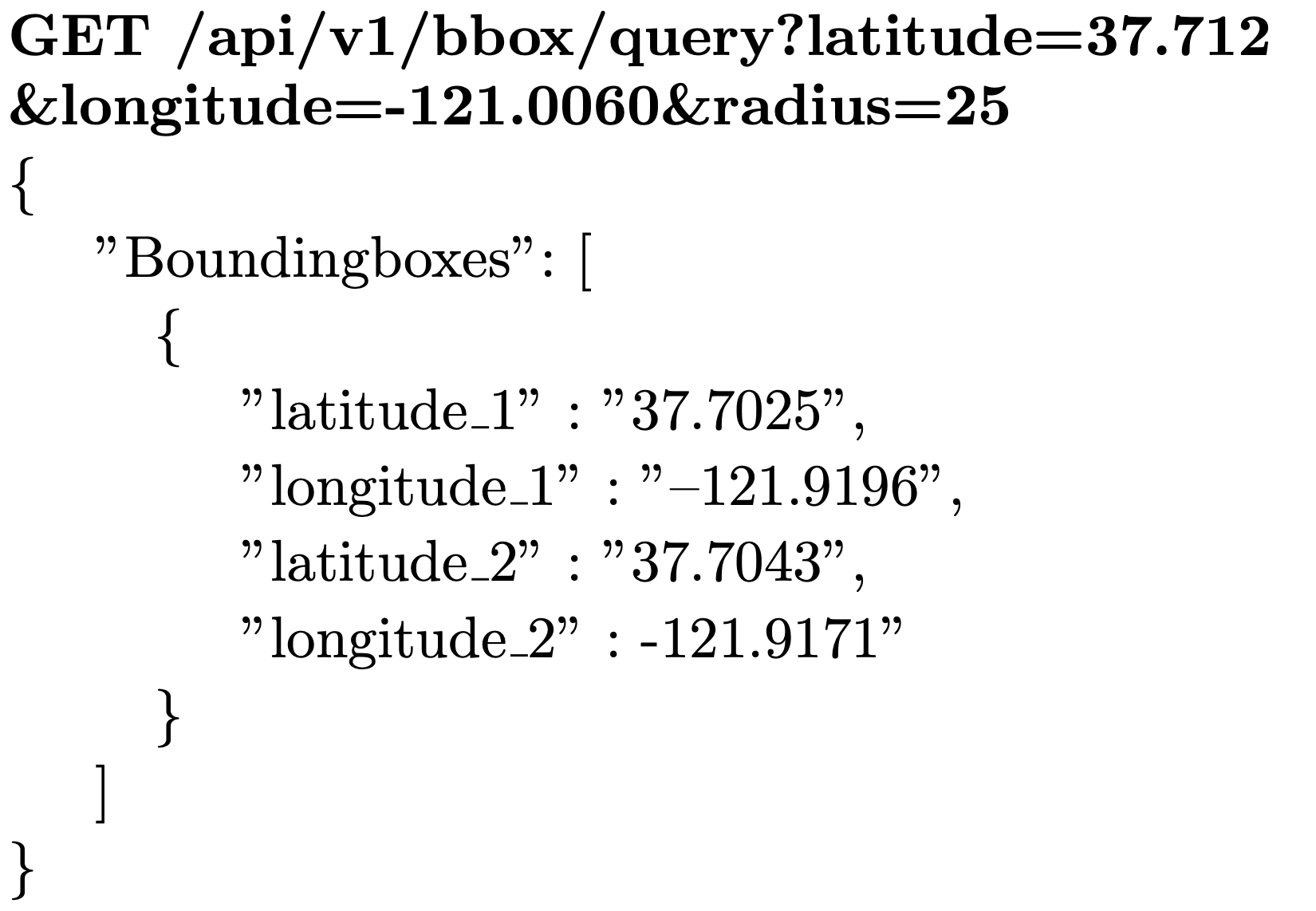}
                \caption{Fetch bounding boxes in a particular radius of the device}
                \label{fig:fetch_bboxes_radius}
            \end{figure}
            
            Fig. \ref{fig:fetch_bboxes_radius} shows the workflow when the device fetches bounding boxes in a particular radius. The Raspberry Pi hosting the database and the API service was connected to a portable battery unit and linked with the device over the device's hotspot, over Wifi. We launched the application when the device was about 50 miles away from any recorded bounding boxes. The application downloaded all the bounding boxes for the country we were testing. We added a few new bounding boxes to the database and started traveling toward the registered bounding boxes. The application kept checking the device's location and attempted to refresh its cache of restricted bounding boxes. When the device came into the 25-mile radius surrounding the recorded bounding boxes, the newer bounding boxes were downloaded to the device cache. When the device was within a restricted bounding box, the camera was prevented from capturing an image. The camera could capture an image when the device was outside all the restricted bounding boxes.

        \end{enumerate}

    \subsection{Storage and performance characteristics of the solution}
        \subsubsection{Storage}~\\
        Our test data consisted of 1 million restricted bounding boxes spread over a large area of a 50-mile radius. The data took up approximately 150 megabytes of data on the database. Initially, we configured the device to only request bounding boxes within a 25-mile radius and noted the memory consumption. When we set up the application to request all restricted bounding boxes in a 50-mile vicinity, the available device memory went down by 120 megabytes.

        \subsubsection{API performance when adding restricted bounding boxes}
        \begin{table}
            \caption{Latency to add a new bounding box with existing bounding boxes in the database}
            \label{tab:lat_new_bbox}
            \begin{tabular}{|p{4cm}|p{1.45cm}|p{1.45cm}|p{1.45cm}|p{1.45cm}|p{1.45cm}|} \hline
                \textbf{\begin{tabular}[c]{@{}l@{}}Num existing \\ bboxes(in 1000s)\end{tabular}} & \textbf{1} & \textbf{100} & \textbf{250} & \textbf{500} & \textbf{1,000} \\ \hline
                \textbf{\begin{tabular}[c]{@{}l@{}}Time to add\\ new bbox(in ms)\end{tabular}}    & 12         & 64          & 139          & 442         & 1695           \\ \hline \end{tabular}
        \end{table}~\\
        As shown in table \ref{tab:lat_new_bbox}, the performance of adding new bounding boxes to the database declined in direct proportion to the number of restricted bounding boxes already in the database. The relationship may be ascribed to the fact that the processing engine needs to check the potential overlap of the new bounding box against existing bounding boxes.

        \subsubsection{API performance when fetching restricted bounding boxes}
        \begin{table}
            \caption{Latency in fetching restricted bounding boxes in current vicinity}
            \label{tab:lat_fetch_bbox}
            \begin{tabular}{|p{4cm}|p{1.45cm}|p{1.45cm}|p{1.45cm}|p{1.45cm}|p{1.45cm}|}
                \hline
                \textbf{\begin{tabular}[c]{@{}l@{}}Num existing \\ bboxes(in 1000s)\end{tabular}}          & \textbf{1} & \textbf{100} & \textbf{250} & \textbf{500} & \textbf{1,000} \\ \hline
                \textbf{\begin{tabular}[c]{@{}l@{}}Time to fetch bboxes\\ in vicinity(in ms)\end{tabular}} & 2          & 15           & 75          & 102          & 851           \\ \hline
            \end{tabular}
        \end{table}~\\
        As shown in Table \ref{tab:lat_fetch_bbox}, the performance of fetching new bounding boxes based on distance from a particular geographical location is better than the performance of adding new bounding boxes. The improved performance  stems from the processing engine's attempts to optimize performance by computing and recording the centroid of each bounding box in the record and quickly computing the distance between the current device location and the centroid of each bounding box.\\

        \subsubsection{Performance of the application when it needs to check whether the device is currently within a bounding box}
        \begin{table}
            \caption{Application startup latency metrics}
            \label{tab:startup_lat}
            \begin{tabular}{|p{4cm}|p{1.45cm}|p{1.45cm}|p{1.45cm}|p{1.45cm}|p{1.45cm}|}
                \hline
                \textbf{\begin{tabular}[c]{@{}l@{}}Num existing \\ bboxes(in 1000s)\end{tabular}}     & \textbf{1} & \textbf{100} & \textbf{250} & \textbf{500} & \textbf{1,000} \\ \hline
                \textbf{\begin{tabular}[c]{@{}l@{}}Application startup\\ latency(in ms)\end{tabular}} & 2          & 50           & 276          & 702          & 2003           \\ \hline
            \end{tabular}
        \end{table}~\\
        As shown in Table \ref{tab:startup_lat}, we see that the additional time taken by the application to check whether the device was currently within a restricted bounding box added modestly to the startup latency. However, in most cases, the additional latency is not prohibitive.

        We attempted to test the performance of the camera application in an area that had 1 million restricted bounding boxes within a 50-mile radius. The performance was tested by measuring the application startup latency. Before the application displays the controls to capture an image, it checks whether the current device location lies within any bounding boxes in its local cache. The performance is similar to the API version when it receives a request to fetch bounding boxes.

\section{Limitations of our approach}
In this section, we discuss the limitations of our approach.
\begin{enumerate}
    \item Our approach largely depends on device manufacturers' voluntary protocol adoption. Therefore, device manufacturers may be unwilling to build devices that conform to these standards. So it may be possible to find devices that can capture photographs of restricted areas.
    \item This approach will only work with devices that have GPS capabilities and periodically active network connections.
    \item GPS data may be spotty or unavailable in areas with tree cover or tall buildings. These factors could "fool" the module to allow photography where it should have been restricted or vice-versa.
    \item A data network might be unavailable in the vicinity of the device. In cases where the restricted bounding boxes could not be refreshed, should the device be allowed to capture images or not?
    \item We have considered but not explored a solution to the problem where malicious software could spoof GPS locations and trick the camera application into believing that it is in someplace where it is not.
\end{enumerate}

\section{Conclusion and future work}
In this paper, we have proposed an approach that promotes national security interests by setting up a global database of restricted bounding boxes representing geographical areas where photography is restricted. We have demonstrated the functional feasibility of the solution and discussed the performance implications of adhering to the new protocol. As photographic device technology evolves, governments and entities worldwide find it increasingly difficult to physically enforce the restriction on photographing sensitive sites. The technological improvements include: gaining compact form factors, becoming more unobtrusive, and achieving more energy efficiency and ubiquitousness. In addition, the physical security posture of sensitive installations of national security importance is improved by implementing a lightweight global standard that device manufacturers must follow.

In extensions of this work, we will focus on the following areas:
\begin{enumerate}
    \item Explore the use of databases geared towards storing geographical data. The  specific database functionality will allow bounding boxes to be sorted and enable geographical computations like distance calculation to be more efficient. The improved performance, in turn, will make the process of adding new bounding boxes, fetching bounding boxes in the vicinity, and checking for location within bounding boxes all very efficient. We propose using the PostGIS extension to the Postgres database to support querying restricted bounding boxes for a specific city or region. For example, suppose the user knows that they are traveling to a particular geographic area. In that case, this feature will enable them to download restricted bounding boxes ahead of time, thereby preventing a lock-out of the photography application or costly downloads over the mobile internet. Geographic databases also allow us to create database indexes on columns containing geographical data. Using indexes will improve the efficiency of geo-SQL queries that run natively on the database.
    \item In today's mobile application, we perform a less-than-efficient process to check whether the device's current location falls within any restricted bounding boxes in the device cache. In future iterations of this work, we plan to use the SpatiaLite extension to \textbf{S}tructured \textbf{Q}uery \textbf{L}anguage (SQL) to allow geographic queries to be run natively in the database. We expect this modification to significantly improve the efficiency of the application, especially during the check to see if the device's current location falls within a restricted bounding box.
    \item We also plan to explore how the same approach may be extended to protect the privacy of individual properties and other geographical locations that could benefit from restricting photography.
\end{enumerate}

%
%


\begin{thebibliography}{6}
%
    \bibitem{dhs:op:readiness} https://www.dhs.gov/publication/operational-readiness-order-photography-and-videotaping-federal-facilities
    
    \bibitem{18:uscode:75} https://www.law.cornell.edu/uscode/text/18/795
    
    \bibitem{18:uscode:797} https://www.law.cornell.edu/uscode/text/50/797
    
    \bibitem {valente:junia:keerthi}
    Valente, J., Koneru, K., Cardenas, A.:Privacy, and Security in Internet-Connected Cameras. In:2019 IEEE International Congress on Internet of Things (ICIOT), pp. 173?180. doi:10.1109/ICIOT.2019.00037

    \bibitem {xiong:cai:han:alrawais}
    Xiong, Z., Cai, Z., Han, Q, and Alrawais, A., Li, W.:ADGAN: Protect Your Location Privacy in Camera Data of Auto-Driving Vehicles. In:IEEE Transactions on Industrial Informatics, pp. 6200?6210.  doi:10.1109/TII.2020.3032352

    \bibitem {liranzo:hayajneh}
    Liranzo, J., Hayajneh, T.:Security and privacy issues affecting cloud-based IP camera. In:2017 IEEE 8th Annual Ubiquitous Computing, Electronics, and Mobile Communication Conference (UEMCON), pp:458?465. doi:10.1109/UEMCON.2017.8249043
    
    \bibitem{yu:chen:wu:cai:cui}
    Yu, J., Chen, H., Wu, K., Cai, Z., Cui, J.:A Distributed Storage System for Robust, Privacy-Preserving Surveillance Cameras. In:2020 IEEE 40th International Conference on Distributed Computing Systems (ICDCS). pp:1195?1196. doi:10.1109/ICDCS47774.2020.00189

    \bibitem{hassan:sazonov}
    Hassan, M., Sazonov, E.:Selective Content Removal for Egocentric Wearable Camera in Nutritional Studies. In:IEEE Access(2020). pp:198615?198623. doi:10.1109/ACCESS.2020.3030723
    
    \bibitem{jiayu:phd:2019}
    Shu, J.:Building intelligent mobile camera systems: visual privacy by design meets social interaction. Hong Kong University of Science and Technology(2019)

    \bibitem{peters:hanvey:veluru:mady:boubekeur:nuseibeh}
    Peters, F., Hanvey, S., Veluru, S., Mady, A., Boubekeur, M., Nuseibeh, B.:Generating Privacy Zones in Smart Cities. In:2018 IEEE International Smart Cities Conference (ISC2). pp:1?8. doi:10.1109/ISC2.2018.8656830

    \bibitem{koufogiannis:pappas}
    Koufogiannis, F., Pappas, G.:Location-dependent privacy. In:2016 IEEE 55th Conference on Decision and Control (CDC). pp:7586?7591. doi:10.1109/CDC.2016.7799441

\end{thebibliography}
\end{document}